\documentclass{emulateapj}
\usepackage{natbib,lscape}
\newcommand{\hii}{\textrm{H}~\textsc{ii}}
\newcommand{\ha}{\textrm{H}\ensuremath{\alpha}}
\newcommand{\hb}{\textrm{H}\ensuremath{\beta}}
\newcommand{\nii}{[\textrm{N}~\textsc{ii}]}
\newcommand{\sii}{[\textrm{S}~\textsc{ii}]}
\newcommand{\oii}{[\textrm{O}~\textsc{ii}]}
\newcommand{\oiii}{[\textrm{O}~\textsc{iii}]}

\newcommand{\hblam}{\textrm{H}\ensuremath{\beta~\lambda4861}}
\newcommand{\oiilam}{\oii~\ensuremath{\lambda3727}}
\newcommand{\oiiilam}{[\textrm{O}~\textsc{iii}]~\ensuremath{\lambda5007}}
\newcommand{\niilam}{[\textrm{N}~\textsc{ii}]~\ensuremath{\lambda6584}} 
\newcommand{\oiiidoublet}{[\textrm{O}~\textsc{iii}]~\ensuremath{\lambda\lambda4959,5007}}
\newcommand{\pagel}{\ensuremath{R_{23}}}
\newcommand{\sfrsb}{\ensuremath{\Sigma(\textrm{SFR})}}
\newcommand{\hasb}{\ensuremath{\Sigma(\ha)}}
\newcommand{\hasbunits}{\textrm{erg~s\ensuremath{^{-1}}~pc\ensuremath{^{-2}}}}
\newcommand{\fluxunits}{\textrm{erg~s\ensuremath{^{-1}}~cm\ensuremath{^{-2}}}}
\newcommand{\mb}{\ensuremath{M_{B}}}

\newenvironment{inlinefigure}{
\def\@captype{figure}
\noindent\begin{minipage}{0.999\linewidth}\begin{center}}
{\end{center}\end{minipage}\smallskip}
\slugcomment{ApJ, in press}
\shorttitle{Integrated Abundances of Disks}
\shortauthors{Moustakas \& Kennicutt}

\begin{document}

\title{Integrated Nebular Abundances of Disk Galaxies} 

\author{John Moustakas\altaffilmark{1} \& Robert~C. Kennicutt,
  Jr.\altaffilmark{1,2}} 
\altaffiltext{1}{Steward Observatory, University of Arizona, 933 N
  Cherry Ave., Tucson, AZ 85721, USA;
  \mbox{jmoustakas@as.arizona.edu}} 
\altaffiltext{2}{Institute of Astronomy, University of Cambridge,
  Madingley Road, Cambridge CB3 0HA, UK; \mbox{robk@ast.cam.ac.uk}} 

\setcounter{footnote}{2}

\begin{abstract}
We study whether integrated optical spectroscopy of a disk galaxy can
be used to infer the mean, or characteristic gas-phase oxygen
abundance in the presence of systematic effects such as spatial
abundance variations, contributions to the integrated emission-line
spectrum from diffuse-ionized gas, and dust attenuation.  Our sample
consists of $14$ nearby disk galaxies with integrated
spectrophotometry, and observations of more than $250$ individual
\hii{} regions culled from the literature.  We consider both
theoretical and empirical strong-line abundance calibrations based on
the $R_{23}\equiv(\oii+\oiii)/\hb$ parameter.  We find that the
integrated oxygen abundance correlates well with the gas-phase
abundance measured at a fixed galactocentric radius, as determined by
the \hii-region abundance gradient.  The typical scatter in the
correlation is $\pm0.1$~dex, independent of the abundance calibration,
or whether the observed integrated emission-line fluxes, the
reddening-corrected fluxes, or the emission-line equivalent widths are
used.  Integrated abundances based on the observed fluxes or
equivalent widths, however, are susceptible to additional systematic
effects of order $0.05-0.1$~dex, at least for the range of reddenings
and stellar populations spanned by our sample.  Unlike the integrated
$R_{23}$ parameter, we find that the integrated \nii/\ha{} and
\sii/\ha{} ratios are enhanced with respect to line-ratios typical of
\hii{} regions, consistent with a modest contribution from
diffuse-ionized gas emission.  We conclude that the $R_{23}$ parameter
can be used to reliably measure the gas-phase abundances of distant
star-forming galaxies.
\end{abstract}



\section{INTRODUCTION}\label{sec:intro}

Improving constraints on the chemical evolution histories of
star-forming galaxies requires abundance measurements with
well-understood systematic uncertainties.  Quantifying these
uncertainties is particularly important in light of recent results
which show that typical star-forming galaxies undergo a relatively
modest amount of chemical enrichment since $z\sim1$
\citep[J.~Moustakas et~al. 2006, in preparation;][]{lilly03, maier04,
kobulnicky03b, kobulnicky04, savaglio05}.  

In the nearby universe, galaxies are resolved into individual \hii{}
regions, enabling a direct measurement of the gas-phase abundance at
discreet spatial positions.  By comparison, spectra of distant
galaxies are a luminosity-weighted average of star-forming regions
spanning a range of physical conditions, stars, and diffuse gas.  In
this paper we investigate whether spatially unresolved spectroscopy of
spiral galaxies encompassing all, or a significant fraction of the
light, reliably measures the gas-phase abundance as determined from
spectroscopy of individual \hii{} regions.

Our analysis focuses on disk galaxies for two reasons.  First,
luminous star-forming disk galaxies at high redshift are more likely
to be observed compared to low-luminosity dwarf galaxies.  Second,
several physical properties of disk galaxies may systematically bias
abundance measurements based on integrated spectroscopy.  For example,
whereas spiral galaxies exhibit radial abundance variations, to first
order, dwarf galaxies are chemically homogenous
\citep[e.g.,][]{kobulnicky96}.  Therefore, we anticipate the largest
discrepancy between integrated and spatially resolved abundances to
arise in a sample of disk galaxies.

One of the most striking and characteristic properties of disk
galaxies is the tendency for their centers to be more metal rich than
their outskirts.  Although the exact physical processes that give rise
to radial abundance gradients are still a matter of debate \citep[and
references therein]{molla96}, it has been shown that the vast majority
of nearby disk galaxies, if not all, possess radial gradients
\citep[e.g.,][]{vila-costas92, zaritsky94, vanzee98}.  Since distant
galaxies may have steeper abundance gradients, indicative of their
earlier evolutionary state \citep[e.g.,][]{molla97}, it is important
to quantify how variations in the slope of the abundance gradient
affects the abundances we infer from integrated spectroscopy.

Another common, if not ubiquitous physical property of spiral
galaxies, including the Milky Way, is the diffuse, low
surface-brightness phase of the interstellar medium called the
diffuse-ionized gas (DIG), or the diffuse-ionized medium
\citep[e.g.,][]{reynolds90, haffner99, ferguson96, wang97,
greenawalt98, zurita00, thilker02}.  Of order $30-50\%$ of the
integrated \ha{} luminosity in spiral galaxies can be attributed to
the DIG, making it an energetically important component of the
interstellar medium of disk galaxies \citep[and references
therein]{calzetti04}.  Since low-ionization line-ratios such as
\nii/\ha{} and \sii/\ha{} in the DIG are enhanced relative to ratios
involving high-excitation lines such as \oiii/\hb{}
\citep[e.g.,][]{martin97, hoopes03}, we want to determine what effect,
if any, the integrated DIG emission of a disk galaxy has on the
abundance we measure from its integrated spectrum.

Finally, it is important to explore how variations in dust attenuation
might bias integrated abundance measurements.  Since the \ha{} line
becomes inaccessible to ground-based optical spectroscopy above
$z\approx0.4$, it may not always be possible to correct the nebular
lines for dust reddening (e.g., \citealt{kobulnicky03b}, but see,
e.g., \citealt{liang04a} and \citealt{maier05}).  Therefore, we
compare the abundances derived from the observed line fluxes, the
reddening-corrected fluxes, and the emission-line equivalent widths,
which, to first order, are insensitive to the effects of dust
\citep{kobulnicky03a}.  In the presence of all these systematic
effects, we want to determine whether we can measure the
``integrated'' gas-phase abundance of a disk galaxy, and, if so, how
we can physically interpret that measurement.

Previous efforts investigating the integrated abundances of spiral
galaxies found that integrated spectroscopy provides a fairly reliable
measure of the gas-phase abundance \citep{kobulnicky99a, pilyugin04b}.
However, these studies synthesized integrated spectra by co-adding
observations of individual \hii{} regions; therefore, they could not
investigate the systematic effects of DIG emission and dust reddening.
Recently, we have obtained spatially unbiased (integrated)
spectroscopy for several hundred nearby galaxies \citep[hereafter
MK06]{moustakas06a}.  Here, we combine these data with spectroscopy of
individual \hii{} regions culled from the literature to investigate
whether chemical analysis methods can be used to estimate the
metallicities of spatially unresolved star-forming galaxies.


\section{DATA}\label{sec:data} 

\subsection{Sample Selection and Emission-Line
  Measurements}\label{sec:sample}  

Investigating the integrated nebular abundances of disk galaxies
quantitatively poses several observational challenges.  First, because
of practical considerations, \hii-region abundance measurements are
generally only available for nearby galaxies subtending relatively
large ($\gtrsim2\arcmin$) angular diameters on the sky.  At the same
time, obtaining an integrated spectrum of an object that is larger
than the typical length of a long-slit ($3\arcmin-5\arcmin$) can be
very difficult.  Nevertheless, our survey (MK06) includes among the
first integrated spectra of spiral galaxies with measurable radial
abundance gradients.

As part of a larger effort to study extra-galactic star formation, we
have obtained integrated, optical ($3600-6900$~\AA) spectrophotometry
at $\sim8$~\AA{} FWHM resolution for a diverse sample of $417$ nearby
($<150$~Mpc) galaxies (MK06).  Our survey implements the
\citet{kenn92a} long-slit drift-scanning technique to obtain
integrated spectra of large galaxies at the spectral resolution
afforded by a $2\farcs5\times200\arcsec$ long-slit \citep[see
also][]{jansen00a, gavazzi04}.  The size of each galaxy at the
$B_{25}$~mag~arcsec$^{-2}$ isophote dictates the parameters of the
drift-scan, and the width of the extraction aperture.  The resulting
spectra typically include $80-100\%$ of the optical light.  These
observations are ideally suited to our analysis because they are
analogous to obtaining a normal (spatially fixed) optical spectrum of
a distant galaxy, which subtends a corresponding smaller angular
diameter on the sky.

In MK06 we present the fluxes and equivalent widths (EWs) of the
strong nebular emission lines, including \oiilam, \oiiilam, \niilam,
\ha, and \hb, for the complete sample of galaxies.  Hereafter, we
assume that \oiii~$\lambda4959$ is $0.34$ times the \oiiilam{}
intensity, as determined by atomic physics \citep{storey00}.
Unfortunately, the temperature-sensitive \oiii~$\lambda4363$ line is
typically not detected in our spectra, which is not surprising given
that this line becomes vanishingly weak in metal-rich galaxies
\citep[e.g.,][]{kenn03}; therefore, we must rely on so-called
strong-line calibrations to estimate the oxygen abundance (see
\S\ref{sec:intoh}).

The nebular line measurements presented in MK06 include a careful
treatment of underlying stellar absorption, which is among the most
important systematic effects to consider in any emission-line
abundance analysis.  In spectroscopic studies of individual \hii{}
regions, the Balmer lines are typically corrected for a constant
$\sim2$~\AA{} of stellar absorption, corresponding to the amount of
absorption expected for a young ($\lesssim20$~Myr) stellar population
\citep[e.g.,][]{mccall85, olofsson95}.  However, in general, the
integrated, luminosity-weighted stellar population of a galaxy will
differ considerably from that of a single star-forming region.  In
particular, the assumption of a constant amount of stellar absorption
for \ha{} and \hb{} may break down, since different stellar
populations will dominate the integrated light at red ($\sim6500$~\AA)
and blue ($\sim4800$~\AA) wavelengths, according to the particular
star-formation history of each galaxy.  Moreover, the exact correction
depends on spectral resolution, since the nebular lines are typically
narrower than the underlying absorption.

To address these issues, in MK06 we use population synthesis to model
the integrated absorption-line spectrum of each galaxy in our sample.
We fit a non-negative linear combination of seven instantaneous-burst
\citet{bruzual03} population synthesis models, convolved to the
instrumental resolution, with ages ranging from $5$~Myr to $12$~Gyr.
Including a simple treatment of dust reddening, this method results in
precise subtraction of the stellar continuum underlying the nebular
emission lines \citep[see also][]{ho97, tremonti04, cid05}.  For the
subset of galaxies considered in this analysis (see below), the \hb{}
absorption ranges from $2.8$ to $4.9$~\AA, with a mean value of
$3.9\pm0.5$~\AA.  The corresponding correction for \ha{} is
$1.8-3.0$~\AA, with a mean of $2.5\pm0.3$~\AA.

We define our sample according to two selection criteria.  First, we
restrict the objects observed by MK06 to spiral galaxies with
well-detected ($3\sigma$) \ha{} and \hb{} emission lines.  We also
reject galaxies whose integrated emission-line spectrum is clearly
dominated by an active nucleus (e.g., NGC~1068).  Next, we conduct a
search of the literature for spectroscopy of individual \hii{} regions
in the remaining objects.\footnote{We only consider observations that
include the \oii, \oiii, and \hb{} nebular lines, since these are the
most likely lines to be measured in distant galaxies.  This
requirement eliminates the imaging spectrophotometric data of the
galaxies in our sample presented by \citet{dutil99}, \citet{martin92},
and \citet{martinbelley97}, among others.} Our search results in data
on $252$ \hii{} regions in $14$ disk galaxies.  The median number of
\hii{} regions per galaxy is $15$.  One object, NGC~4713, has fewer
than six \hii{} regions, which \citet{zaritsky94} advocate as the
minimum number of regions needed to robustly determine the abundance
gradient \citep[see also][]{dutil01}.  Nevertheless, we retain
NGC~4713 in our sample, bearing in mind that the derived abundance
gradient may be subject to additional systematic uncertainty.  To
place the size of our sample in context, these $14$ objects account
for $\sim1/4$ of the total number of galaxies with measurable
abundance gradients \citep[e.g.,][]{pilyugin04a}.

In Table~\ref{table:intoh} we list the galaxies in our sample and the
global properties relevant to our analysis.  Column (2) gives the
morphological type of each object from \citet[hereafter RC3]{devac91}.
Column (3) lists the absolute $B$-band magnitude, based on the
Galactic extinction-corrected $B$-band magnitude
\citep[$R_{V}=3.1$;][]{schlegel98, odonnell94} and distance tabulated
in MK06.  Column (4) lists $\rho_{25}$, the radius of the semi-major
axis at the $B_{25}$ mag~arcsec$^{-2}$ isophote from the RC3.  In
columns (5) and (6), respectively, we list the photometric inclination
angle and position angle, as determined in the $K_{s}$-band
\citep{jarrett00, jarrett03}.  To compute the inclination angle we use
the observed major-to-minor axis ratio at the $K_{s,
20}$~mag~arcsec$^{-2}$ isophote \citep{jarrett00, jarrett03}, and
assume that the axial ratio of a system viewed edge-on is $0.2$
\citep{tully98}.  For NGC~5194, however, we use the kinematic
inclination angle determined by \citet{tully74}.  Finally, in column
(7) we give the \ha{} surface brightness,
$\hasb=4.5\times10^{44}I(\ha)/\rho_{25}^{2}$ in \hasbunits, where
$I(\ha)$ is the \ha{} flux in \fluxunits{} from MK06, corrected for
extinction as described in \S\ref{sec:intoh}.  The remaining columns
in Table~\ref{table:intoh} are described in \S\ref{sec:intoh}.

For each \hii{} region drawn from the literature, we tabulate the
coordinates relative to the galactic nucleus, and the stellar
absorption and reddening-corrected \oiilam, \oiiilam, and \hb{}
emission-line flux measurements and uncertainties.  As before, we set
$\oiii~\lambda4959 = 0.34\times\oiiilam$.  Using the optical disk
radius and galactic position and inclination angles given in
Table~\ref{table:intoh}, we also compute the de-projected
galactocentric radius of each \hii{} region, normalized to
$\rho_{25}$.  In \S\ref{sec:hiioh} we use these data to determine the
radial abundance gradients for the galaxies in our sample.

\subsection{Sample Properties and Selection Biases}\label{sec:biases}   

In this section we discuss some of the sample selection biases.  As
alluded to in \S\ref{sec:sample}, the size of the sample is determined
by simultaneously requiring integrated spectroscopy \emph{and}
spectroscopy for a sufficient number of individual \hii{} regions.  In
terms of $B$-band luminosity, the sample is generally biased toward
bright galaxies, since lower-luminosity galaxies tend not to exhibit
abundance gradients.  We note, however, that our sample spans more
than an order-of-magnitude in $B$-band luminosity, from $\mb=-18.2$ to
$-21.0$~mag.  For comparison, the absolute $B$ magnitudes of the disk
galaxies studied by \citet{pilyugin04a}, which is the most
comprehensive compilation of galaxies with measured abundance
gradients to date, range from $-17.8$ to $-21.6$~mag.  The
distribution of inclination angles in our sample is fairly broad,
ranging from $11^{\circ}$ in NGC~3344, to $72^{\circ}$ in NGC~3198;
the mean inclination angle is $41^{\circ}\pm19^{\circ}$.  In
\S\ref{sec:results} we test whether variations in inclination along
the line-of-sight systematically biases the integrated abundances. 

The distribution of morphological types in our sample is weighted
toward late-type galaxies: $85\%$ of the sample is type Sbc or later,
with only two early-type objects, NGC~3351 (Sb) and NGC~4736 (Sab).
This bias arises because of observational and physical selection
effects which tend to prefer late-type galaxies.  For example, the
number of early-type galaxies with measured abundance gradients is
relatively limited, because \hii{} regions in early-type galaxies are
rarer and have lower luminosities and surface brightnesses than
corresponding \hii{} regions in late-type galaxies
\citep[e.g.,][]{kenn88}.  Previous studies have found that early-type
galaxies are, on average, more metal rich, and have shallower
abundance gradients than late-type galaxies \citep[e.g.,][]{garnett87,
oey93, zaritsky94, dutil99}.  We note, however, that our sample
includes galaxies that are as luminous and metal-rich as other
well-known early-type disk galaxies such as M~81 \citep{garnett87}.  A
related selection effect is that in the integrated spectra of
early-type galaxies the equivalent widths of the nebular lines, if
present, tend to be significantly lower than the corresponding
emission-line equivalent widths in late-type galaxies.  This
observation can be attributed to the lower star-formation rate per
unit luminosity or mass in early-type galaxies
\citep[e.g.,][]{caldwell91, kenn94}.

Despite the relatively small number of early-type galaxies, we argue
that our sample is still representative for the kind of analysis we
are conducting, and that our conclusions are not strongly affected by
this morphological bias.  In particular, the objects most likely to be
studied at high redshift should have higher gas resevoirs, high
specific star-formation rates, and large emission-line equivalent
widths.  For example, in an analysis of $64$ star-forming galaxies at
$0.3<z<0.8$, \citet{kobulnicky03b} found an average \hb{}
emission-line EW of $21\pm14$~\AA, while \citet{pettini01} measured
\hb{} EWs in excess of $\sim20$~\AA{} in a sample of five Lyman-break
galaxies at $z\sim3$.  Finally, since early-type galaxies tend to have
shallow abundance gradients, small amounts of dust, and a weak or
absent DIG phase, we do not expect the integrated abundances of
early-type galaxies to differ significantly from the abundances
infered from spectroscopy of individual \hii{} regions.

\section{ANALYSIS}\label{analysis}\label{sec:analysis}

\subsection{Integrated Abundances}\label{sec:intoh}

We use the emission-line measurements from MK06 to compute the
integrated oxygen abundances of the galaxies selected in
\S\ref{sec:sample}.  A principal assumption of this method is that the
observed emission lines arise via photoionization from massive stars.
Consequently, it is important to assess any possible contamination
from nuclear activity in our integrated emission-line spectra.
\citet{ho97} have obtained nuclear spectra of all the galaxies in our
sample; they find that $50\%$ ($7/14$) have \hii-like nuclei, $36\%$
($5/14$) are LINERs or transition objects, and $14\%$ ($2/14$) exhibit
Seyfert~2 nuclear line-ratios.  In all cases, however, we find that
the nucleus contributes $<3\%$ to the integrated \ha{} flux.
Therefore, we can apply standard techniques to derive chemical
abundances from the observed emission-line spectra.

From the observed \ha/\hb{} ratio, we estimate the nebular reddening
assuming an intrinsic ratio of $2.86$ and the \citet{odonnell94} Milky
Way extinction curve (listed in col. [8] of Table~\ref{table:intoh}).
The distribution of reddening values for the sample ranges from $0.08$
to $0.69$~mag, with a mean value of $0.31\pm0.17$~mag.
\citet{moustakas06b} show that adopting a simple Milky Way extinction
law works remarkably well at accounting for the effects of dust on the
nebular lines from integrated optical spectra \citep[see
also][]{kewley02b}, despite the wide variation in attenuation laws
predicted by theoretical models \citep[e.g.,][]{witt00, fischera05}.

To investigate the systematic effect of dust attenuation, we
inter-compare the abundances determined using the observed
(un-corrected) emission-line fluxes, the reddening-corrected fluxes,
and the emission-line EWs.  \citet{kobulnicky03a} have suggested
replacing fluxes with EWs when determining the integrated
emission-line abundances of galaxies, since, to first order, EWs are
insensitive to dust attenuation.  However, the EW method for deriving
abundances is sensitive to intrinsic variations in the stellar
populations, and to differential extinction between the lines and the
continuum, as we discuss in \S\ref{sec:results}.

We explore the effect of choosing a particular strong-line calibration
by computing abundances using both the \citet[hereafter
M91]{mcgaugh91} and the \citet[hereafter PT05]{pilyugin05}
calibrations of the $\pagel\equiv(\oiilam + \oiiidoublet)/\hb$
parameter.  We focus on the \pagel{} parameter as an abundance
diagnostic because the oxygen and \hb{} lines are among the strongest
optical emission lines, and they are observable in ground-based
optical spectra to $z\sim1$.  Both calibrations account for variations
in ionization parameter at fixed metallicity, either through the
\oiii/\oii{} ratio (M91), or the \oiii/(\oii+\oiii) ratio (PT05).  The
M91 calibration, as parameterized by \citet{kobulnicky99a}, is based
on an extensive grid of photoionization models spanning a wide range
of metallicity and ionization parameter.  By comparison, PT05 have
constructed a purely empirical calibration, based on a carefully
selected sample of \hii{} regions with well-measured electron
temperatures.  Finally, we note that although \pagel{} is
double-valued as a function of abundance (see, e.g., M91), all the
galaxies in our sample lie on the upper branch of the relation between
\pagel{} and (O/H).

On average, the M91 abundance scale is shifted to higher values
relative to the PT05 scale by $\sim0.5$~dex, or a factor of $\sim3$
(see Fig.~\ref{fig:slopes}).  In \S\ref{sec:results} we show that
selecting among the M91 and PT05 calibrations has a significant effect
on the absolute oxygen abundances, and a modest effect on the slope of
the derived abundance gradients, but does not significantly alter our
conclusions.  Using a different theoretical calibration in place of
M91 \citep[e.g.,][]{zaritsky94, kewley02a, kobulnicky04} has a
negligible effect on our conclusions.

\begin{inlinefigure}
\begin{center}
\plotone{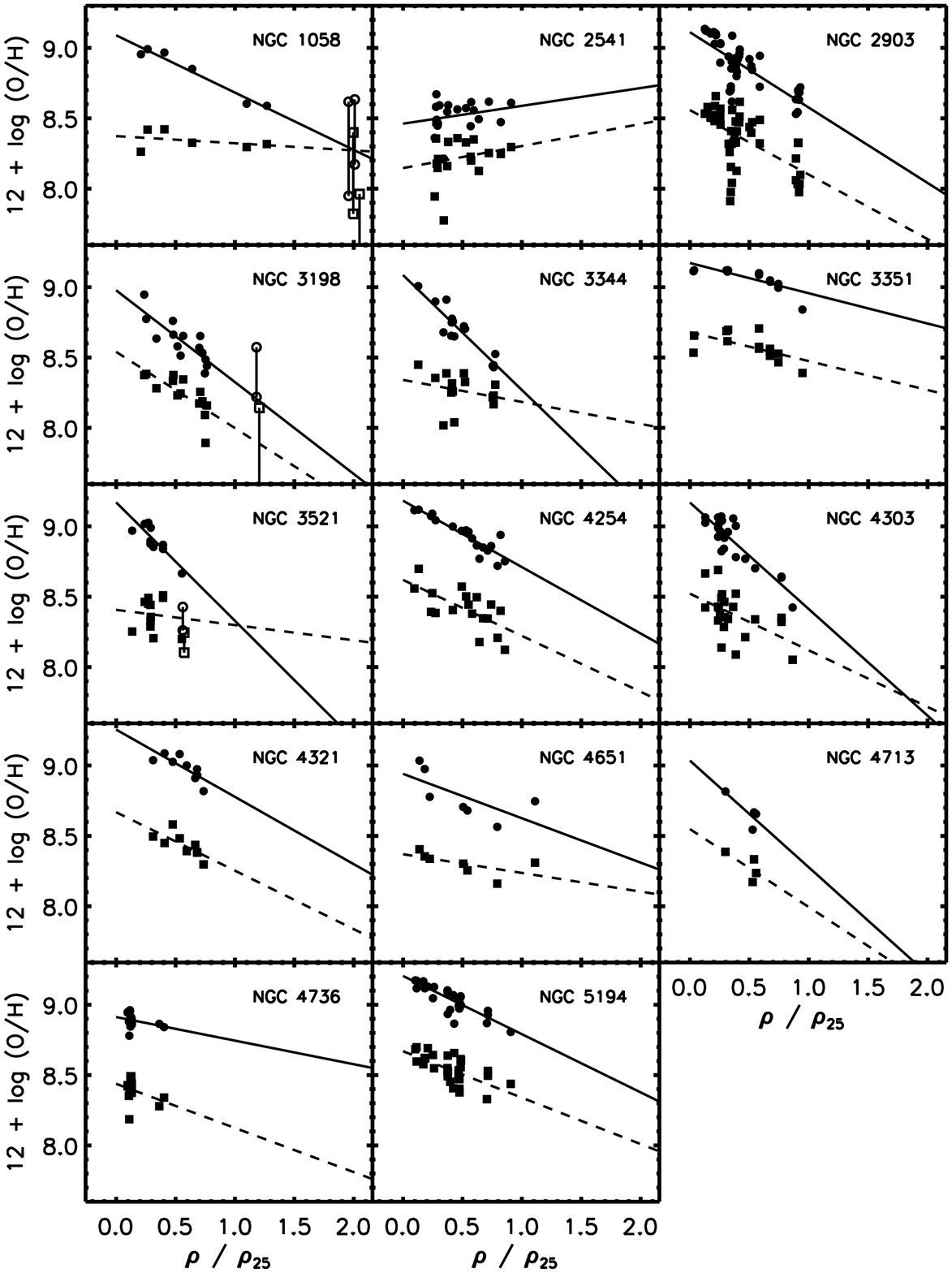}
\figcaption{Oxygen abundance vs.~de-projected galactocentric radius,
$\rho$, normalized to $\rho_{25}$, the disk radius at the
$B_{25}$~mag~arcsec$^{-2}$ isophote.  Oxygen abundances for each
\hii{} region have been computed using both the \citet[M91]{mcgaugh91}
({\em circles}) and the \citet[PT05]{pilyugin05} ({\em squares})
abundance calibrations.  The solid line is the best fit through the
M91 abundance distribution, and the dashed line is the corresponding
fit through the PT05 abundances.  \hii{} regions that have been
excluded from the fits are shown as open symbols, with a vertical line
connecting the upper- and lower-branch abundances (see
\S\ref{sec:hiioh} for details).
\label{fig:gradients}} 
\end{center}
\end{inlinefigure}

In Columns (9)--(11) of Table~\ref{table:intoh} we list the three
integrated abundances and statistical uncertainties derived using the
M91 abundance calibration; columns (12)--(14) give the corresponding
abundances based on the PT05 calibration.  In every object except
NGC~3351 and NGC~4321, the two dustiest galaxies in the sample, we
detect the \oii, \oiii, and \hb{} emission lines with $>3\sigma$
significance.  In NGC~3351 and NGC~4321, however, we derive an upper
limit to the \oii{} flux and EW, which translates into a lower limit
on the oxygen abundance.

\subsection{H~{\sc ii}-Region Abundance Gradients}\label{sec:hiioh}

We compute the radial abundance gradient of each galaxy in our sample
using an unweighted linear least-squares fit to both the M91 and the
PT05 \hii-region abundances.  To quantify the uncertainty in the
gradient, we re-compute the fit $500$ times, modulating the
metallicity of each \hii{} region by a Gaussian distribution with a
width equal to the statistical uncertainty.  Table~\ref{table:hiioh}
and Figure~\ref{fig:gradients} present the results of our fits.
Columns (2)--(4) of Table~\ref{table:hiioh} are based on the M91
abundance calibration and list, respectively, the extrapolated central
abundance at galactocentric radius $\rho=0$, the \emph{characteristic}
abundance, defined as the oxygen abundance at $\rho=0.4\,\rho_{25}$
\citep{zaritsky94, garnett02}, and the slope of the abundance gradient
in ${\rm dex}~\rho_{25}^{-1}$.  Columns (5)--(7) give the
corresponding quantities derived using the PT05 abundance calibration.
Finally, columns (8) and (9) give the number of \hii{} regions used in
the fit, and the corresponding references to the literature,
respectively.

In Figure~\ref{fig:gradients} we plot the normalized galactocentric
radius, $\rho/\rho_{25}$, versus oxygen abundance for the \hii{}
regions in each galaxy.  The circles and squares distinguish the M91
and PT05 abundances, respectively.  We assume that all the \hii{}
regions lie on the upper branch of the appropriate \pagel-(O/H)
relation, except as noted below.  The solid line is the best fit
through the M91 abundance distribution, and the dashed line is the
corresponding fit through the PT05 \hii-region abundances.

We exclude a handful of \hii{} regions from the fits, shown in
Figure~\ref{fig:gradients} as open symbols without error bars, with a
solid vertical line connecting the upper- and lower-branch \pagel{}
abundances.  Specifically, in NGC~1058 we exclude the two outermost
\hii{} regions, FGW1058G and FGW1058H \citep{ferguson98}, because they
are well outside the main optical disk of the galaxy.  We also exclude
the outermost \hii{} region in NGC~3198, --027--187
\citep{zaritsky94}, which has an extremely low ionization parameter,
$\log\,(\oiii/\oii)=-0.76$.  Unfortunately, the strong-line
calibrations used here fail at reproducing the ionization properties
of this object, which leads to the unphysical situation of having a
lower-branch abundance that is several times higher than the
corresponding upper-branch abundance.  Finally, we also reject the
anomolously low-metallicity \hii{} region in NGC~3521, --033--118
\citep{zaritsky94}, which has a negligible effect on the derived
abundance gradient.

\begin{inlinefigure}
\begin{center}
\plotone{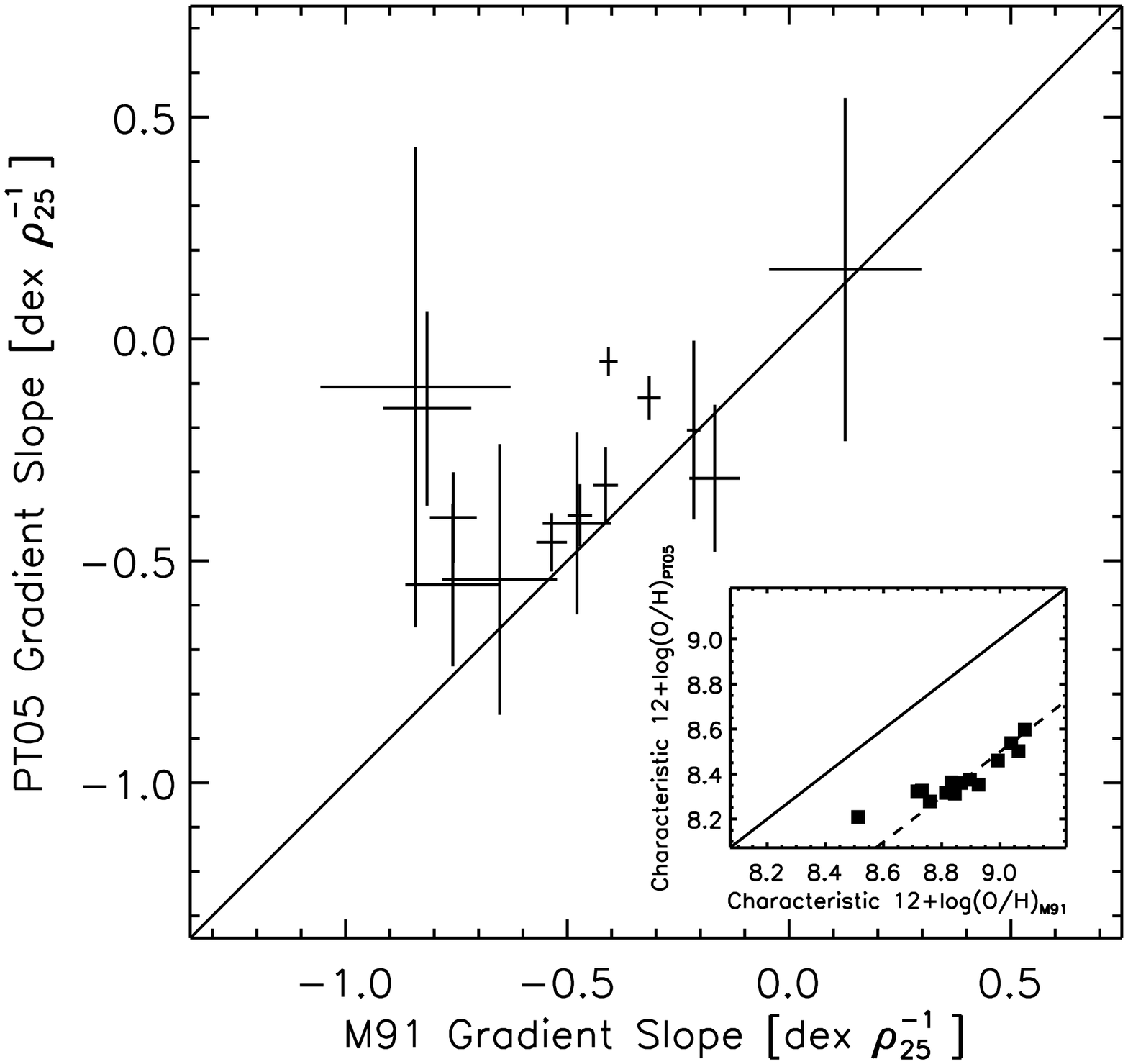}
\figcaption{Slope of the radial abundance gradient determined from
fitting to the \citet[PT05]{pilyugin05} abundance distribution vs.~the
corresponding slope derived from fitting to the \citet[M91]{mcgaugh91}
\hii-region abundances (see Fig.~\ref{fig:gradients}).  (\emph{Inset})
Correlation between the \emph{characteristic} oxygen abundances (see
\S\ref{sec:hiioh} and Table~\ref{table:hiioh} for details).  The
dashed line shows the $0.5$~dex mean offset between the M91 and PT05
abundance scales, while the solid line in both panels is the
line-of-equality.
\label{fig:slopes}}  
\end{center}
\end{inlinefigure}

In Figure~\ref{fig:slopes} we compare the slope of the gradients
derived using the M91 and PT05 abundance calibrations, and a
comparison of the characteristic abundances (\emph{inset}).  The solid
line in both panels is the line-of-equality, and the dashed line in
the inset shows the median $0.5$~dex offset between the M91 and PT05
abundances.  The two slopes generally correlate, although the PT05
gradients are systematically shallower than the corresponding
gradients derived using the M91 oxygen abundances.  The four most
discrepant examples are NGC~3521, NGC~3344, NGC~1058, and NGC~4303.
In all cases, the PT05 calibration predicts a much flatter
relationship between galactocentric position and oxygen abundance, a
result which has profound implications for our understanding of the
chemical evolution properties of spiral galaxies
\citep[e.g.,][]{molla96, pilyugin06}.  Exploring this result in more
detail, however, is beyond the scope of this paper.

\section{RESULTS}\label{sec:results}

Figures~\ref{fig:ohvsoh_m91} and \ref{fig:ohvsoh_pt05} illustrate the
principal result of this paper.  Here we plot the characteristic
oxygen abundance, as derived from the \hii-region abundance gradient
(Fig.~\ref{fig:gradients}, Table~\ref{table:hiioh}), versus our three
estimates of the integrated abundance.  We plot the integrated
``observed'', ``reddening-corrected'', and ``EW'' abundances using
open circles, filled circles, and open squares, respectively, and the
lower limits to the integrated abundances in NGC~3351 and NGC~4321 as
arrows (\S\ref{sec:intoh} and Table~\ref{table:intoh}).

As an independent check on our results, in Figure~\ref{fig:ohvsoh_m91}
we show the integrated abundances determined by \citet{kobulnicky99a},
against the characteristic \hii-region abundances from
\citet{zaritsky94} ({\em filled grey triangles}).
\citet{kobulnicky99a} derive pseudo-integrated abundances for their
sample of $22$ galaxies, nine of which are common to our sample, by
dividing each object into $5-13$ equal-size radial zones and
convolving the observed \ha{} surface-brightness distribution with the
reddening-corrected \hii-region line-strengths from
\citet{zaritsky94}.  We subtract $0.2$~dex from these points to place
them on the M91 abundance scale.

\begin{inlinefigure}
\begin{center}
\plotone{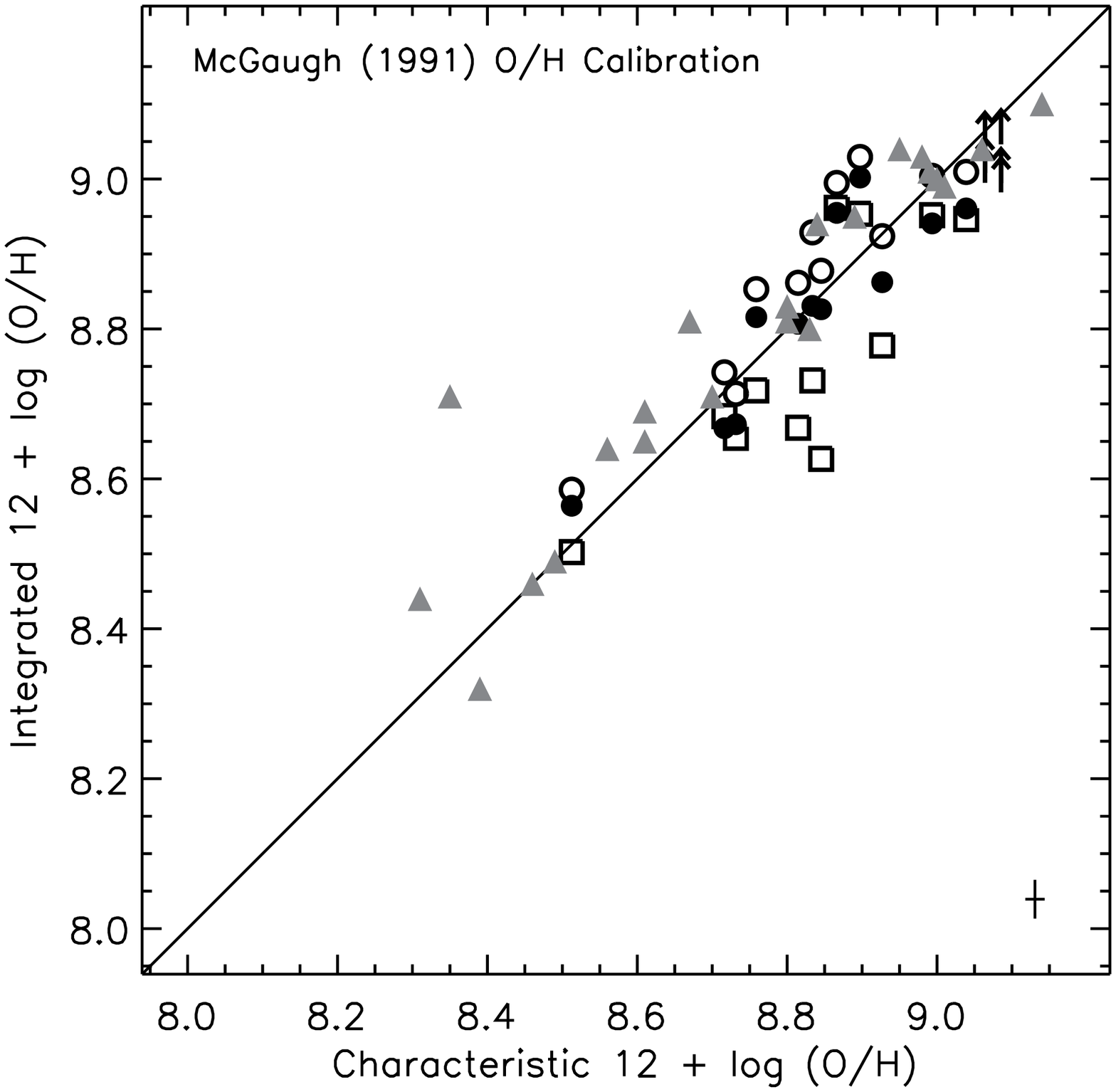}
\figcaption{Characteristic vs.~integrated oxygen abundance, based on the
\citet{mcgaugh91} abundance calibration.  Symbols differentiate
between abundances computed using the observed emission-line fluxes
({\em open circles}), the reddening-corrected fluxes ({\em filled
circles}), and the emission-line equivalent widths ({\em open
squares}).  For comparison we plot the results from
\citet{kobulnicky99a} using filled grey triangles, accounting for the
$0.2$~dex offset between the \citet{zaritsky94} and \citet{mcgaugh91}
abundance scales.  Lower limits to the oxygen abundance in NGC~3351
and NGC~4321 are shown as arrows.  The cross shows the median
measurement uncertainty of the data.
\label{fig:ohvsoh_m91}}
\end{center}
\end{inlinefigure}

Examining Figures~\ref{fig:ohvsoh_m91} and \ref{fig:ohvsoh_pt05}, we
find that all three integrated abundances correlate with the
characteristic, or spatially resolved abundance, independent of the
abundance calibration used.  A Spearman rank correlation test
quantifies the significance of the observed correlations at $>99\%$.
We conclude, therefore, that, to first order, the abundance inferred
from the integrated emission-line spectrum of a galaxy is
representative of the gas-phase oxygen abundance at
$\rho=0.4\,\rho_{25}$, even in the presence of an abundance gradient,
contributions from DIG emission, and variations in dust reddening.

\begin{inlinefigure}
\begin{center}
\plotone{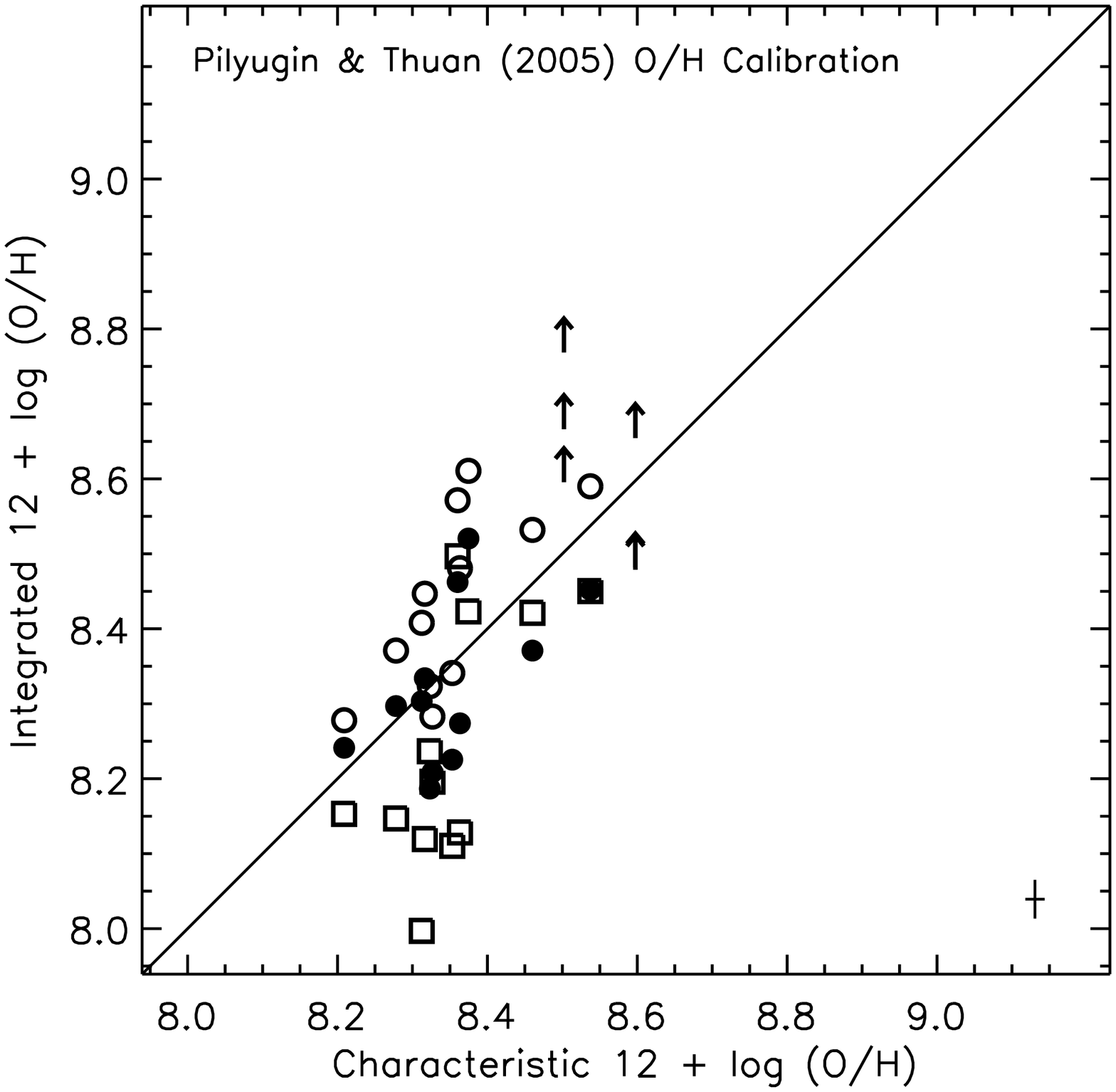}
\figcaption{Same as Fig.~\ref{fig:ohvsoh_m91}, but using the
\citet{pilyugin05} abundance calibration.\label{fig:ohvsoh_pt05}} 
\end{center}
\end{inlinefigure}

Table~\ref{table:resid} presents the mean, standard deviation, and
median of the residuals we measure between the integrated and
characteristic abundances.  On average, the integrated abundance
inferred from the observed (un-corrected) line-fluxes overestimates
the characteristic abundance by $0.05\pm0.06$~dex or $0.09\pm0.08$~dex
using the M91 or PT05 calibration, respectively.  Accounting for dust
reddening using a simple Milky Way extinction curve \citep{odonnell94}
removes the systematic offset, but has almost no effect on the
residual scatter.  This result is consistent with other recent
studies, which find that the integrated emission lines of nearby
star-forming galaxies appear to be attenuated by a simple foreground
dust screen, and that the integrated Balmer decrement is a robust
indicator of the amount of dust \citep[e.g.,][]{calzetti94, jansen01,
kewley02b, dopita02, moustakas06b}.  The relatively small wavelength
separation between \oiilam{} and \hblam, and the broad similarity of
extinction and attenuation curves in the optical, additionally help to
minimize the effects of dust.

The integrated abundances determined from the emission-line EWs, by
comparison, systematically \emph{underestimate} the characteristic
abundance.  The mean offset is $-0.06\pm0.09$~dex using the M91
calibration, or $-0.11\pm0.13$~dex using the PT05 calibration.  The
median offsets are slightly smaller, $-0.04$~dex and $-0.09$~dex,
respectively.  \citet{kobulnicky03a} discuss in detail the systematic
effects of using EWs to determine integrated abundances.  In
particular, following their recommendation, we have assumed that
$\alpha=1$, where $\alpha$ is related to the ratio of the intrinsic
continuum between $3727$~\AA{} and $4861$~\AA, and the differential
extinction between the stellar continuum and the emission lines (see
\citealt{kobulnicky03a}, their eq. [7]).  Assuming $\alpha=0.9$, which
is well within the expected range of this parameter
\citep[$\alpha=0.84\pm0.3$;][]{kobulnicky03a}, increases the M91
integrated EW abundances by a median $+0.04$~dex, and the
corresponding PT05 abundances by $+0.06$~dex.  We conclude, therefore,
that the integrated EW abundances are consistent with the
characteristic gas-phase abundances, provided that an appropriate
choice for $\alpha$ is made for the sample under consideration.  We
attribute the increased scatter in the integrated EW abundances
relative to the abundances based on the line-fluxes to variations in
$\alpha$ among star-forming galaxies.  We note, however, that the
measured scatter is comparable to the $\pm0.1$~dex precision of the EW
abundance technique \citep{kobulnicky03a}.

\begin{inlinefigure}
\begin{center}
\plotone{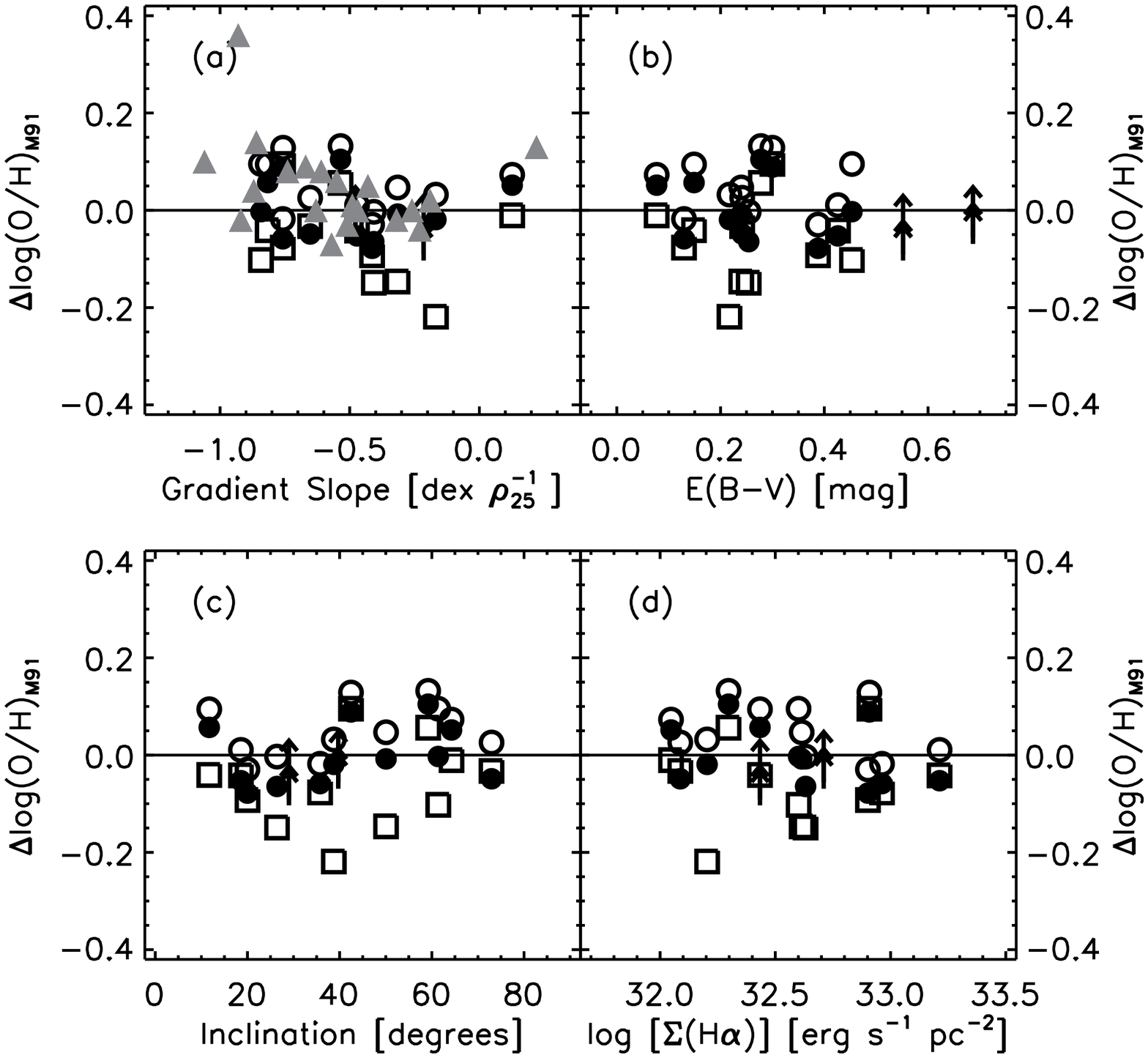}
\figcaption{Oxygen abundance residuals (integrated minus characteristic
abundance) based on the \citet{mcgaugh91} calibration vs.~(\emph{a})
the slope of the abundance gradient, (\emph{b}) dust reddening,
(\emph{c}) inclination angle, and (\emph{d}) \ha{} surface brightness.
Symbols are as in Fig.~\ref{fig:ohvsoh_m91}.
\label{fig:ohresid_m91}}
\end{center}
\end{inlinefigure}

\begin{inlinefigure}
\begin{center}
\plotone{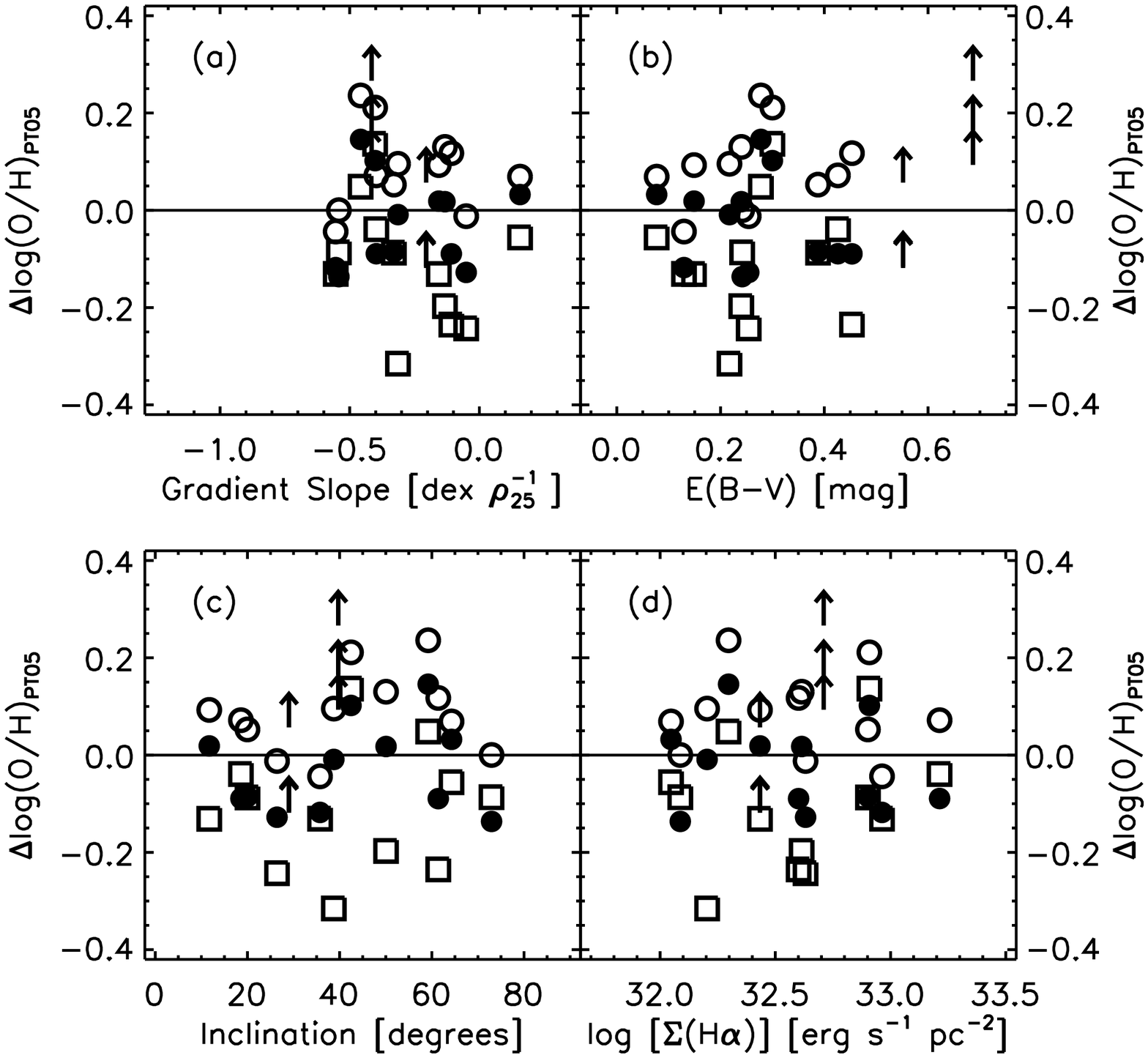}
\figcaption{Same as Fig.~\ref{fig:ohresid_m91}, but using the
\citet{pilyugin05} abundance calibration.  \label{fig:ohresid_pt05}} 
\end{center}
\end{inlinefigure}

Next, we investigate the physical origin of the residual scatter
between the integrated and characteristic abundances.  In
Figure~\ref{fig:ohresid_m91} we plot the abundance residuals based on
the M91 calibration, versus various physical properties of the sample:
(\emph{a}) the slope of the abundance gradient, (\emph{b}) dust
reddening, (\emph{c}) inclination angle, and (\emph{d}) \ha{} surface
brightness, \hasb, which characterizes the disk star-formation rate
(SFR) per unit area \citep{kenn98b}.  For our sample \hasb{} spans a
factor of $\sim15$, comparable to the range in $\hasb\propto\sfrsb$
found by \citet{kenn98b} for a larger sample of normal disk galaxies.
The corresponding residual plots using the PT05 abundance calibration
are shown in Figure~\ref{fig:ohresid_pt05}.
Table~\ref{table:spearman} presents the Spearman rank correlation
coefficients for each residual plot, and the probability that the
variables are \emph{uncorrelated}.

Examining Figures~\ref{fig:ohresid_m91}--\ref{fig:ohresid_pt05} and
Table~\ref{table:spearman}, we find that the residuals are not highly
correlated with \emph{any} of the variables we have considered.  We
find a marginally significant ($\geq80\%$ probability) correlation
between the observed and reddening-corrected M91 abundance residuals,
and the galaxy inclination angle (Fig.~\ref{fig:ohresid_m91}\emph{c}).
We also find a marginally significant anti-correlation between these
residuals and \hasb{} (Fig.~\ref{fig:ohresid_m91}\emph{d}).  None of
the residual correlations using the PT05 calibration are significant,
although the increased scatter in the PT05 abundances
(Fig.~\ref{fig:gradients}, Table~\ref{table:resid}) may be masking any
underlying trends.

The positive correlation between the abundance residuals and
inclination angle, if it exists, is opposite of what we would have
expected.  The correlation indicates that the integrated abundance of
an edge-on galaxy is \emph{higher} than the corresponding metallicity
of a less-inclined galaxy.  Given the existence of radial abundance
gradients (e.g., Fig.~\ref{fig:gradients}), we would have anticipated
the integrated spectrum of an inclined galaxy to be weighted more
toward the outer, metal-poor \hii{} regions.  This exact trend is
shown by \citet[their Fig.~7]{tremonti04} as a strong correlation
between the residuals of the stellar-mass metallicity relation and
inclination angle, based on an analysis of $\sim50,000$ star-forming
galaxies in the Sloan Digital Sky Survey.  \citet{tremonti04} find
that edge-on galaxies may appear, on average, $0.2$~dex ($60\%$) more
metal-rich than face-on galaxies.  Since this systematic difference is
comparable to the amount of chemical enrichment between the present
day and $z\sim1$ \citep[e.g., J.~Moustakas et~al. 2006, in
preparation;][]{kobulnicky03b, kobulnicky04}, sample selection effects
that may be biased with respect to inclination angle should be
carefully considered in look-back studies of galactic chemical
evolution \citep[e.g., Tully-Fisher studies;][]{mouhcine06a}.

\begin{figure}
\begin{center}
\includegraphics[scale=0.35]{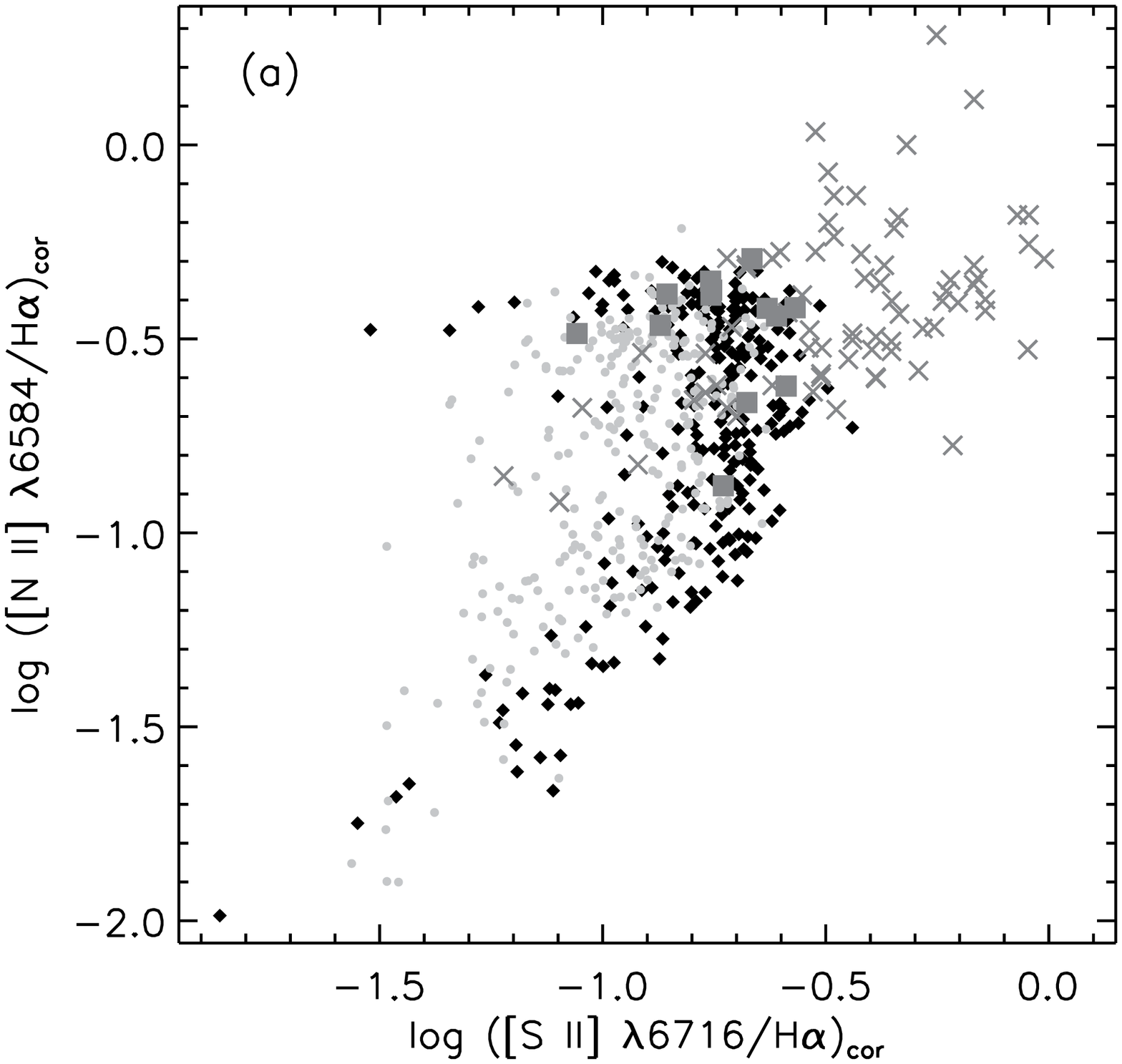}
\includegraphics[scale=0.35]{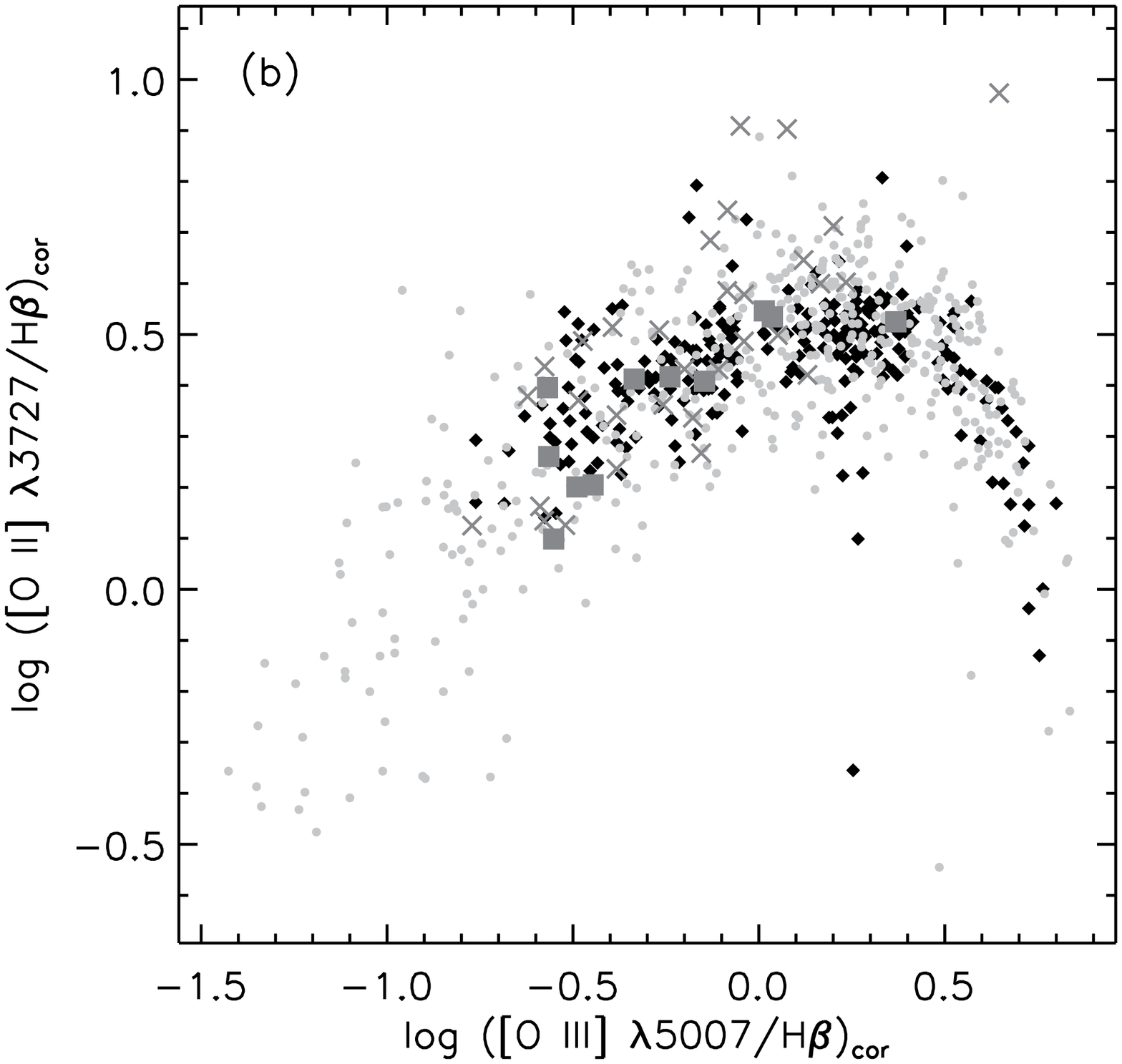}
\figcaption{Diagnostic diagrams showing the emission-line sequences
traced by our sample ({\em squares}), the full sample of star-forming
galaxies from \citet{moustakas06a} ({\em diamonds}), \hii{} regions
({\em grey points}), and regions of diffuse-ionized gas in M~33,
M~51a, and M~31 ({\em crosses}), as described in \S\ref{sec:results}.
(\emph{a}) Reddening-corrected \sii~$\lambda6716$/\ha{} ratio
vs.~\niilam/\ha.  (\emph{b}) Reddening-corrected \oiiilam/\hb{} ratio
vs.~\oiilam/\hb.
\label{fig:ratios}}
\end{center}
\end{figure}

Since the strength of low-ionization line-ratios such as \nii/\ha{}
and \sii/\ha{} correlate with the \ha{} surface brightness
\citep[e.g.,][]{wang98}, indicating a prominent DIG phase, the lack of
a significant correlation between \hasb{} and the abundance residuals
in Figures~\ref{fig:ohresid_m91}\emph{d} and
\ref{fig:ohresid_pt05}\emph{d} suggests that integrated abundance
measurements are relatively unaffected by DIG emission.  To explore
this result in more detail, in Figure~\ref{fig:ratios} we plot two
emission-line diagnostic diagrams for our sample ({\em squares}), all
the star-forming galaxies with integrated spectroscopy in the MK06
survey ({\em diamonds}), individual \hii{} regions drawn from a
diverse sample of disk and dwarf galaxies \citep[{\em grey
points};][]{mccall85, zaritsky94, vanzee98, izotov98}, and regions of
DIG in three disk galaxies ({\em crosses}), NGC~0598=M~33,
NGC~5194=M~51a, and NGC~0223=M~31, based on deep long-slit
spectroscopy by \citet{galarza99} and \citet{hoopes03}.

Figure~\ref{fig:ratios}\emph{a} shows that \hii{} regions and galaxies
trace out similar emission-line sequences in the \nii/\ha{} versus
\sii/\ha{} plane (corrected for reddening), while the DIG regions
extend to much higher values of both \nii/\ha{} and \sii/\ha.
Moreover, we observe that the integrated line-ratios preferentially
lie at or above the upper envelope defined by the \hii{} regions
\citep[see also][]{lehnert94}.  The enhancement of low-ionization
lines in the DIG is generally attributed to a low ionization parameter
\citep[e.g.,][]{domgorgen94, martin97}.  Alternatively, recent
self-consistent theoretical models of aging star-forming regions
reveal that old, low surface-brightness \hii{} regions exhibit higher
\nii/\ha{} and \sii/\ha{} ratios relative to young, luminous \hii{}
regions (M.~Dopita 2006, private communication).  Regardless of its
physical origin, our integrated spectra show a clear enhancement of
low-ionization emission relative to \hii{} regions.  One implication
of this result is that \nii- or \sii-based strong-line abundance
diagnostics applied to integrated spectra of galaxies
\citep[e.g.,][]{charlot01, denicolo02, pettini04} may be moderately
biased toward higher metallicities.

In Figure~\ref{fig:ratios}\emph{b} we plot the reddening-corrected
\oiii/\hb{} ratio versus \oii/\hb.  In this diagram the emission-line
sequences defined by galaxies, \hii{} regions, and regions of DIG
generally overlap.  Unlike \nii/\ha{} and \sii/\ha, lines of constant
ionization parameter run nearly parallel to the \oiii/\hb{} versus
\oii/\hb{} emission-line sequence \citep[e.g.,][]{kewley02a}.  We
conclude, therefore, that measurements based on the globally averaged
\pagel{} parameter give abundances that are fairly representative of
the gas-phase oxygen abundances of galaxies.

We conclude this section by testing one of the principal assumptions
of the preceding analysis.  Following \citet{zaritsky94}, we have
defined the characteristic abundance of a disk galaxy as the oxygen
abundance at $\rho=0.4\,\rho_{25}$ \citep[see also][]{garnett97,
garnett02}.  However, given the form of the abundance gradient, we can
\emph{derive} the normalized galactocentric radius, $\rho_{\rm
int}/\rho_{25}$, where the integrated abundance equals the gas-phase
abundance, assuming that the reddening-corrected integrated abundance
is a perfect measure of the gas-phase abundance.  For this calculation
we only consider the M91 abundance calibration, which produces smaller
uncertainties in the gradient slopes; we also exclude NGC~2541, which
exhibits an increasing abundance with radius (although the slope is
formally consistent with zero).

We find that $\rho_{\rm int}/\rho_{25}$ ranges from $0.2$ in NGC~2903,
to $0.8$ in NGC~3344, with a mean value of $0.47\pm0.15$.  Not
surprisingly, $\rho_{\rm int}/\rho_{25}$ correlates with the slope of
the abundance gradient, in the sense that the integrated abundances of
galaxies with steeper abundance gradients are systematically higher
than the integrated abundances of galaxies with shallower gradients.
We conclude, therefore, that the dominant source of scatter in the
abundance residuals found in
Figures~\ref{fig:ohresid_m91}--\ref{fig:ohresid_pt05} (see also
Table~\ref{table:resid}) is due to assuming a fixed
$\rho_{int}/\rho_{25}=0.4$.  Our analysis shows that while this is a
reasonable average assumption, the exact radius does vary from
object-to-object in a way that depends mildly on the slope of the
abundance gradient.

\section{DISCUSSION \& SUMMARY}\label{sec:conclusions} 

In this paper we have investigated whether spatially unresolved
(integrated) spectroscopy of a spiral galaxy can be used to infer its
mean interstellar oxygen abundance, as measured from observations of
individual \hii{} regions.  Our analysis makes use of a new integrated
spectrophotometric survey of nearby star-forming galaxies presented by
\citet{moustakas06a}, and observations of more than $250$ \hii{}
regions culled from the literature.  Our final sample consists of
$14$, predominantly late-type spiral galaxies spanning a wide range of
$B$-band luminosity, inclination angle, \ha{} surface brightness, and
dust content (Table~\ref{table:intoh}).  For the first time, these
data allow us to investigate the systematic effects of diffuse-ionized
gas emission, dust reddening, and the slope of the abundance gradient
on the determination of integrated abundances.

We derive integrated oxygen abundances using the emission-line
measurements presented by \citet{moustakas06a}, who carry out a
careful treatment of underlying stellar absorption.  We apply both a
theoretical \citep{mcgaugh91}, and an empirical \citep{pilyugin05}
strong-line (\pagel) abundance calibration to the observed
emission-line fluxes, the reddening-corrected fluxes, and the
emission-line equivalent widths \citep[following][]{kobulnicky03a}.
We analyze the \hii-region data compiled from the literature
self-consistently; in particular, we determine the \hii-region
abundances using the same abundance calibration applied to the
integrated emission lines, and we compute the de-projected
galactocentric position of each object using robust inclination and
position angles determined in the near-infrared.  With these
measurements, we derive the form of the abundance gradient for each
galaxy in our sample (Fig.~\ref{fig:gradients},
Table~\ref{table:hiioh}).

Our principal result is that the integrated abundance of a normal disk
galaxy correlates well with its \emph{characteristic} gas-phase
abundance, or the \hii-region abundance measured at
$\rho=0.4\,\rho_{25}$ (Figs.~\ref{fig:ohvsoh_m91} and
\ref{fig:ohvsoh_pt05}).  The typical scatter in the correlation is
$\pm0.1$~dex, independent of whether the \citet{mcgaugh91} or the
\citet{pilyugin05} calibration is used (Table~\ref{table:resid}).  We
note, however, that the scatter in the abundance residuals is
generally higher when using the Pilyugin \& Thuan calibration.  These
results confirm and extend the analysis carried out by
\citet{kobulnicky99a} and \citet{pilyugin04b}, who used simulated
integrated galaxy spectra.  Integrated abundances based on the
observed emission-line fluxes, or the emission-line equivalent widths
\citep{kobulnicky03a} are susceptible to additional systematic effects
of order $0.05-0.1$~dex with respect to the abundances based on the
reddening-corrected fluxes, at least for the range of reddenings and
stellar populations spanned by our sample.  The systematic residuals
using the equivalent-width method, however, can be minimized or
eliminated by adopting a different value of $\alpha$ \citep[see
\S\ref{sec:results} and][]{kobulnicky03a}.

We studied the residuals between the integrated and characteristic
abundances of the galaxies in our sample to identify a physical origin
for the source of the scatter.  We found no significant correlations
between the residuals and the slope of the abundance gradient, the
dust reddening, the inclination angle, or the \ha{} surface brightness
(Figs.~\ref{fig:ohresid_m91} and \ref{fig:ohresid_pt05}, and
Table~\ref{table:spearman}).  We note, however, that
\citet{tremonti04} found that inclination can significantly bias
integrated abundances, in the sense that edge-on galaxies may appear,
on average, $0.2$~dex more metal-poor than face-on galaxies.  

By comparing the emission-line sequences of galaxies, \hii{} regions,
and regions of diffuse-ionized gas, we found that the integrated
\nii/\ha{} and \sii/\ha{} ratios of galaxies lie along the upper
envelope of the sequence defined by high surface brightness \hii{}
regions, consistent with a modest contribution from diffuse-ionized
gas emission to our integrated spectra (Fig.~\ref{fig:ratios}).  By
comparison, the integrated \oiii/\hb{} and \oii/\hb{} ratios of
galaxies are relatively insensitive to contributions from
diffuse-ionized gas emission, suggesting that the \pagel{} parameter
is a more robust abundance diagnostic for galaxies than strong-line
calibrations that rely on either \sii{} or \nii.

Finally, we found that the reddening-corrected integrated abundances
of the galaxies in our sample were equal to the gas-phase abundance at
normalized galactocentric radii ranging from $0.2$ to $0.8$, depending
on the slope of the abundance gradient.  On average, the two
abundances were equal at $\rho/\rho_{25}=0.47\pm0.15$, indicating that
the commonly adopted radius used to define the characteristic
abundance of a disk galaxy (i.e., $\rho/\rho_{25}=0.4$) is a
reasonable assumption.  

We conclude that the integrated \pagel{} parameter of star-forming
disk galaxies is a robust tracer of the gas-phase oxygen abundance,
even in the presence of a variety of systematic effects.  The strength
of the \hb, \oii, and \oiii{} lines under typical nebular conditions,
and their accessibility to ground-based optical spectrographs to
$z\sim1$, ensures that the \pagel{} parameter will continue to be used
to measure the build-up of heavy elements in the interstellar media of
galaxies with cosmic time.

\acknowledgements

The authors would like to thank Christy Tremonti, Aleks
Diamond-Stanic, and Dennis Zaritsky for comments on a draft of the
paper that helped improve its content and clarity.  We also thank the
anonymous referee for several valuable suggestions that led to a much
improved version of the paper.  Funding for this project has been
provided by NSF grant AST-0307386, NASA grant NAG5-8426, and a SINGS
grant, provided by NASA through JPL contract 1224769.  This research
has made use of the NASA/IPAC Extragalactic Database, which is
operated by the Jet Propulsion Laboratory, California Institute of
Technology, under contract with the National Aeronautics and Space
Administration.




\clearpage

\begin{landscape}
\begin{deluxetable}{llccccccccccccc}
\tabletypesize{\tiny}
\tablecaption{Integrated Oxygen Abundances\label{table:intoh}}
\tablewidth{0pt}
\tablehead{
\multicolumn{2}{c}{} & \colhead{$M_{B}$} & \colhead{$\rho_{25}$} & \colhead{$i$} & \colhead{$\theta$} & \colhead{$\log\,[\Sigma({\rm H}\alpha)]$} & \colhead{$E(B-V)$} & \multicolumn{3}{c}{12+log(O/H)$_{\rm M91}$\tablenotemark{a}} & \colhead{} & \multicolumn{3}{c}{12+log(O/H)$_{\rm PT05}$\tablenotemark{a}} \\ 
\cline{9-11}\cline{13-15}
\colhead{Galaxy} & \colhead{Type} & \colhead{(mag)} & \colhead{(arcmin)} & \colhead{(deg)} & \colhead{(deg)} & \colhead{(erg~s$^{-1}$~pc$^{-2}$)} & \colhead{(mag)} & \colhead{Observed} & \colhead{Corrected} & \colhead{EWs} & \colhead{} & \colhead{Observed} & \colhead{Corrected} & \colhead{EWs} \\ 
\colhead{(1)} & 
\colhead{(2)} & 
\colhead{(3)} & 
\colhead{(4)} & 
\colhead{(5)} & 
\colhead{(6)} & 
\colhead{(7)} & 
\colhead{(8)} & 
\colhead{(9)} & 
\colhead{(10)} & 
\colhead{(11)} & 
\colhead{} & 
\colhead{(12)} & 
\colhead{(13)} & 
\colhead{(14)} 
}
\startdata
NGC 1058 &  SA(rs)c &  -18.4 &  1.51 &  26 &  95 &  32.6 & 
$0.25\pm0.08$ &  $8.92\pm0.03$ &  $8.86\pm0.03$ &  $8.78\pm0.05$ &   & 
$8.34\pm0.04$ &  $8.23\pm0.04$ &  $8.11\pm0.06$ \\ 
NGC 2541 &  SA(s)cd &  -18.2 &  3.15 &  64 &  175 &  32.0 & 
$0.08\pm0.08$ &  $8.59\pm0.04$ &  $8.56\pm0.04$ &  $8.50\pm0.04$ &   & 
$8.28\pm0.04$ &  $8.24\pm0.05$ &  $8.15\pm0.06$ \\ 
NGC 2903 &  SB(s)d &  -20.2 &  6.30 &  59 &  13 &  32.3 & 
$0.28\pm0.08$ &  $9.03\pm0.02$ &  $9.00\pm0.03$ &  $8.95\pm0.04$ &   & 
$8.61\pm0.05$ &  $8.52\pm0.06$ &  $8.42\pm0.08$ \\ 
NGC 3198 &  SB(rs)c &  -19.9 &  4.26 &  72 &  40 &  32.1 & 
$0.24\pm0.09$ &  $8.74\pm0.06$ &  $8.67\pm0.07$ &  $8.68\pm0.06$ &   & 
$8.32\pm0.07$ &  $8.19\pm0.09$ &  $8.24\pm0.08$ \\ 
NGC 3344 &  (R)SAB(r)bc &  -19.2 &  3.54 &  11 &  150 & 
32.4 &  $0.15\pm0.08$ &  $8.85\pm0.03$ &  $8.82\pm0.04$ &  $8.72\pm0.05$ &   & 
$8.37\pm0.05$ &  $8.30\pm0.05$ &  $8.15\pm0.06$ \\ 
NGC 3351 &  SB(r)b &  -19.6 &  3.71 &  28 &  163 &  32.4 & 
$0.55\pm0.11$ &  $>9.05$ &  $>8.99$ &  $>8.98$ &   &  $>8.65$ &  $>8.48$ & 
$>8.48$ \\ 
NGC 3521 &  SAB(rs)bc &  -20.1 &  5.48 &  61 &  166 &  32.6 & 
$0.45\pm0.07$ &  $8.93\pm0.02$ &  $8.83\pm0.04$ &  $8.73\pm0.05$ &   & 
$8.48\pm0.03$ &  $8.27\pm0.04$ &  $8.13\pm0.05$ \\ 
NGC 4254 &  SA(s)c &  -20.3 &  2.69 &  18 &  24 &  33.2 & 
$0.43\pm0.07$ &  $9.00\pm0.01$ &  $8.94\pm0.02$ &  $8.95\pm0.02$ &   & 
$8.53\pm0.02$ &  $8.37\pm0.03$ &  $8.42\pm0.03$ \\ 
NGC 4303 &  SAB(rs)bc &  -20.0 &  3.23 &  42 &  20 & 
32.9 &  $0.30\pm0.07$ &  $8.99\pm0.01$ &  $8.95\pm0.02$ &  $8.96\pm0.02$ &   & 
$8.57\pm0.02$ &  $8.46\pm0.03$ &  $8.50\pm0.03$ \\ 
NGC 4321 &  SAB(s)bc &  -21.0 &  3.71 &  39 &  108 & 
32.7 &  $0.69\pm0.09$ &  $>9.04$ &  $>8.99$ &  $>9.01$ &   &  $>8.77$ & 
$>8.60$ &  $>8.67$ \\ 
NGC 4651 &  SA(rs)c &  -20.3 &  1.99 &  50 &  80 &  32.6 & 
$0.24\pm0.07$ &  $8.86\pm0.02$ &  $8.81\pm0.03$ &  $8.67\pm0.04$ &   & 
$8.45\pm0.03$ &  $8.33\pm0.04$ &  $8.12\pm0.05$ \\ 
NGC 4713 &  SAB(rs)d &  -18.6 &  1.35 &  35 &  50 &  33.0 & 
$0.13\pm0.07$ &  $8.71\pm0.02$ &  $8.67\pm0.03$ &  $8.65\pm0.03$ &   & 
$8.28\pm0.03$ &  $8.21\pm0.04$ &  $8.20\pm0.04$ \\ 
NGC 4736 &  (R)SA(r)ab &  -19.4 &  5.61 &  38 &  85 & 
32.2 &  $0.22\pm0.07$ &  $8.88\pm0.02$ &  $8.83\pm0.02$ &  $8.63\pm0.03$ &   & 
$8.41\pm0.02$ &  $8.30\pm0.03$ &  $8.00\pm0.04$ \\ 
NGC 5194 &  SA(s)bc &  -20.7 &  5.61 &  20 &  58 & 
32.9 &  $0.39\pm0.07$ &  $9.01\pm0.01$ &  $8.96\pm0.02$ &  $8.95\pm0.02$ &   & 
$8.59\pm0.03$ &  $8.45\pm0.03$ &  $8.45\pm0.03$
\enddata
\tablecomments{Col. (1) Galaxy name; Col. (2) Morphological type from \citet{devac91}; Col. (3) Absolute $B$-band magnitude, corrected for foreground Galactic extinction \citep[$R_{V}=3.1$;][]{schlegel98, odonnell94}, based on the apparent magnitude and distance given by \citet{moustakas06a}; Col. (4) Radius of the major axis at the $B_{25}$~mag~arcsec$^{-2}$ isophote from \citet{devac91}; Col. (5) Galaxy inclination angle based on the $K_{s}$-band major-to-minor axis ratio \citep{jarrett00, jarrett03}, as described in \S\ref{sec:sample}, except for NGC~5194, which is from \citet{tully74}; Col. (6) Galaxy position angle, measured positive from North to East, in the $K_{s}$ band \citep{jarrett00, jarrett03}, as described in \S\ref{sec:sample}; Col. (7) Surface. Col. (8) Nebular reddening determined from the observed H$\alpha$/H$\beta$ ratio \citep{moustakas06a}, assuming an intrinsic ratio of $2.86$ and the \citet{odonnell94} Milky Way extinction curve; Oxygen abundances have been computed using the observed line fluxes [Cols. (9) and (12)], the reddening-corrected line fluxes [Cols. (10) and (13)], and the emission-line equivalent widths (EWs) [Cols. (11) and (14)].}
\tablenotetext{a}{Oxygen abundances derived using either the \citet[M91]{mcgaugh91} or the \citet[PT05]{pilyugin05} abundance calibrations.  See \S\ref{sec:intoh} for details.}
\end{deluxetable}
\clearpage
\end{landscape}

\begin{deluxetable}{lccccccccc}
\tabletypesize{\tiny}
\tablecaption{\ion{H}{2}-Region Oxygen Abundance Gradients\label{table:hiioh}}
\tablewidth{0pt}
\tablehead{
\colhead{} & \multicolumn{3}{c}{M91\tablenotemark{a}} & \colhead{} & \multicolumn{3}{c}{PT05\tablenotemark{a}} & \multicolumn{2}{c}{} \\ 
\cline{2-4}\cline{6-8}
\colhead{} & \colhead{12+log(O/H)} & \colhead{12+log(O/H)} & \colhead{Gradient} & \colhead{} & \colhead{12+log(O/H)} & \colhead{12+log(O/H)} & \colhead{Gradient} & \multicolumn{2}{c}{} \\
\colhead{Galaxy} & 
\colhead{at $\rho=0$} & 
\colhead{at $\rho=0.4\,\rho_{25}$} & 
\colhead{(dex $\rho_{25}^{-1}$)} & 
\colhead{} & 
\colhead{at $\rho=0$} & 
\colhead{at $\rho=0.4\,\rho_{25}$} & 
\colhead{(dex $\rho_{25}^{-1}$)} & 
\colhead{N(H~{\sc ii})} & 
\colhead{Refs.} \\ 
\colhead{(1)} & 
\colhead{(2)} & 
\colhead{(3)} & 
\colhead{(4)} & 
\colhead{} & 
\colhead{(5)} & 
\colhead{(6)} & 
\colhead{(7)} & 
\colhead{(8)} & 
\colhead{(9)} 
}
\startdata
NGC 1058 &  $9.09\pm0.01$ &  $8.93\pm0.01$ &  $-0.41\pm0.02$ &   & 
$8.37\pm0.02$ &  $8.35\pm0.01$ &  $-0.05\pm0.03$ &  6 &  1 \\ 
NGC 2541 &  $8.46\pm0.08$ &  $8.51\pm0.04$ &  \phs$0.13\pm0.17$ &   & 
$8.15\pm0.19$ &  $8.21\pm0.08$ &  \phs$0.16\pm0.41$ &  19 &  2 \\ 
NGC 2903 &  $9.11\pm0.01$ &  $8.90\pm0.01$ &  $-0.54\pm0.03$ &   & 
$8.56\pm0.03$ &  $8.37\pm0.01$ &  $-0.46\pm0.07$ &  45 &  2,3,4,5 \\ 
NGC 3198 &  $8.98\pm0.07$ &  $8.72\pm0.03$ &  $-0.65\pm0.13$ &   & 
$8.54\pm0.16$ &  $8.32\pm0.07$ &  $-0.54\pm0.30$ &  14 &  2 \\ 
NGC 3344 &  $9.09\pm0.05$ &  $8.76\pm0.02$ &  $-0.82\pm0.11$ &   & 
$8.34\pm0.10$ &  $8.28\pm0.04$ &  $-0.16\pm0.22$ &  15 &  2,4,6 \\ 
NGC 3351 &  $9.17\pm0.01$ &  $9.09\pm0.01$ &  $-0.21\pm0.02$ &   & 
$8.68\pm0.16$ &  $8.60\pm0.07$ &  $-0.21\pm0.22$ &  14 &  4,7,8 \\ 
NGC 3521 &  $9.17\pm0.06$ &  $8.83\pm0.03$ &  $-0.84\pm0.20$ &   & 
$8.41\pm0.14$ &  $8.36\pm0.07$ &  $-0.11\pm0.50$ &  12 &  2,7 \\ 
NGC 4254 &  $9.18\pm0.01$ &  $8.99\pm0.01$ &  $-0.47\pm0.03$ &   & 
$8.62\pm0.05$ &  $8.46\pm0.02$ &  $-0.40\pm0.07$ &  18 &  4,9,10 \\ 
NGC 4303 &  $9.17\pm0.02$ &  $8.87\pm0.01$ &  $-0.76\pm0.05$ &   & 
$8.52\pm0.05$ &  $8.36\pm0.02$ &  $-0.40\pm0.10$ &  22 &  10,11 \\ 
NGC 4321 &  $9.26\pm0.03$ &  $9.06\pm0.01$ &  $-0.48\pm0.07$ &   & 
$8.67\pm0.11$ &  $8.50\pm0.04$ &  $-0.42\pm0.19$ &  9 &  4,10 \\ 
NGC 4651 &  $8.94\pm0.01$ &  $8.81\pm0.01$ &  $-0.32\pm0.03$ &   & 
$8.37\pm0.04$ &  $8.32\pm0.02$ &  $-0.13\pm0.05$ &  7 &  12 \\ 
NGC 4713 &  $9.03\pm0.05$ &  $8.73\pm0.01$ &  $-0.76\pm0.10$ &   & 
$8.55\pm0.08$ &  $8.33\pm0.02$ &  $-0.55\pm0.17$ &  4 &  12 \\ 
NGC 4736 &  $8.91\pm0.01$ &  $8.85\pm0.01$ &  $-0.17\pm0.06$ &   & 
$8.44\pm0.03$ &  $8.31\pm0.04$ &  $-0.31\pm0.17$ &  15 &  4,7,8 \\ 
NGC 5194 &  $9.20\pm0.01$ &  $9.04\pm0.00$ &  $-0.41\pm0.03$ &   & 
$8.67\pm0.04$ &  $8.54\pm0.01$ &  $-0.33\pm0.08$ &  34 &  4,7,13,14,15
\enddata
\tablecomments{Col. (1) Galaxy name; Cols. (2) and (5) Central abundance (at radius $\rho=0$) based on the derived abundance gradient; Cols. (3) and (6) \emph{Characteristic} abundance (at $\rho=0.4\rho_{25}$), based on the derived abundance gradient; Cols. (4) and (7) Slope of the abundance gradient; Col. (8) Number of H~{\sc ii} regions;  Col. (9) H~{\sc ii}-region references.}
\tablenotetext{a}{Oxygen abundances and abundance gradients derived using either the \citet[M91]{mcgaugh91} or the \citet[PT05]{pilyugin05} abundance calibrations.  See \S\ref{sec:intoh} for details.}
\tablerefs{(1) \citet{ferguson98}; (2) \citet{zaritsky94}; (3) \citet{bresolin05}; (4) \citet{mccall85}; (5) \citet{vanzee98}; (6) \citet{vilchez88b}; (7) \citet{bresolin99}; (8) \citet{oey93}; (9) \citet{henry94}; (10) \citet{shields91}; (11) \citet{henry92}; (12) \citet{skillman96}; (13) \citet{bresolin04}; (14) \citet{diaz91}; (15) \citet{garnett04}.}
\end{deluxetable}

\begin{deluxetable}{lccccc}
\tabletypesize{\small}
\tablecaption{Integrated vs. Characteristic Abundance Residuals\label{table:resid}}
\tablewidth{0pt}
\tablehead{
\colhead{} & \multicolumn{2}{c}{M91\tablenotemark{b}} & \colhead{} & \multicolumn{2}{c}{PT05\tablenotemark{b}} \\ 
\cline{2-3}\cline{5-6}
\colhead{Method\tablenotemark{a}} & 
\colhead{$\langle\Delta\log({\rm O/H})\rangle$} & 
\colhead{$\Delta\log({\rm O/H})_{\rm median}$} & 
\colhead{} & 
\colhead{$\langle\Delta\log({\rm O/H})\rangle$} & 
\colhead{$\Delta\log({\rm O/H})_{\rm median}$} 
}
\startdata
Observed &  \phs$0.049\pm0.055$ &  \phs$0.047$ &   &  \phs$0.085\pm0.083$ & 
\phs$0.093$ \\ 
Corrected &  $-0.003\pm0.063$ &  $-0.008$ &   &  $-0.028\pm0.093$ & 
$-0.009$ \\ 
EWs &  $-0.063\pm0.088$ &  $-0.042$ &   &  $-0.111\pm0.128$ &  $-0.087$
\enddata
\tablenotetext{a}{Integrated abundances based on the observed emission-line fluxes, the reddening-corrected line-fluxes, or the emission-line equivalent widths (EWs).  See \S\ref{sec:intoh} for details.}
\tablenotetext{b}{Oxygen abundances derived using either the \citet[M91]{mcgaugh91} or the \citet[PT05]{pilyugin05} abundance calibrations.  See \S\ref{sec:intoh} for details.}
\end{deluxetable}

\begin{deluxetable}{lccccccccccccc}
\tabletypesize{\small}
\tablecaption{Residual Correlation Coefficients\label{table:spearman}\tablenotemark{a}}
\tablewidth{0pt}
\tablehead{
\colhead{} & \multicolumn{2}{c}{M91\tablenotemark{b}} & \colhead{} & \multicolumn{2}{c}{PT05\tablenotemark{b}} \\ 
\cline{2-3}\cline{5-6}
\colhead{Variables} & 
\colhead{Coefficient} & 
\colhead{Probability} & 
\colhead{} & 
\colhead{Coefficient} & 
\colhead{Probability} 
}
\startdata
$\Delta\log({\rm O/H})_{\rm obs}$ vs. Slope &  $-0.26$ &  \phs$0.42$ &   & 
$-0.07$ &  \phs$0.83$ \\ 
$\Delta\log({\rm O/H})_{\rm obs}$ vs. $E(B-V)$ &  \phs$0.08$ &  \phs$0.81$ & 
 &  \phs$0.31$ &  \phs$0.33$ \\ 
$\Delta\log({\rm O/H})_{\rm obs}$ vs. $i$ &  \phs$0.42$ &  \phs$0.17$ &   & 
\phs$0.22$ &  \phs$0.50$ \\ 
$\Delta\log({\rm O/H})_{\rm obs}$ vs. $\Sigma({\rm H}\alpha)$ &  $-0.39$ & 
\phs$0.21$ &   &  $-0.14$ &  \phs$0.66$ \\ 
\cline{1-6}
$\Delta\log({\rm O/H})_{\rm cor}$ vs. Slope &  $-0.22$ &  \phs$0.50$ &   & 
\phs$0.16$ &  \phs$0.62$ \\ 
$\Delta\log({\rm O/H})_{\rm cor}$ vs. $E(B-V)$ &  $-0.10$ &  \phs$0.76$ &   & 
$-0.13$ &  \phs$0.70$ \\ 
$\Delta\log({\rm O/H})_{\rm cor}$ vs. $i$ &  \phs$0.37$ &  \phs$0.24$ &   & 
\phs$0.04$ &  \phs$0.90$ \\ 
$\Delta\log({\rm O/H})_{\rm cor}$ vs. $\Sigma({\rm H}\alpha)$ &  $-0.41$ & 
\phs$0.18$ &   &  $-0.20$ &  \phs$0.54$ \\ 
\cline{1-6}
$\Delta\log({\rm O/H})_{\rm EW}$ vs. Slope &  $-0.29$ &  \phs$0.35$ &   & 
$-0.13$ &  \phs$0.68$ \\ 
$\Delta\log({\rm O/H})_{\rm EW}$ vs. $E(B-V)$ &  $-0.02$ &  \phs$0.95$ &   & 
\phs$0.19$ &  \phs$0.56$ \\ 
$\Delta\log({\rm O/H})_{\rm EW}$ vs. $i$ &  \phs$0.27$ &  \phs$0.39$ &   & 
\phs$0.16$ &  \phs$0.62$ \\ 
$\Delta\log({\rm O/H})_{\rm EW}$ vs. $\Sigma({\rm H}\alpha)$ &  $-0.11$ & 
\phs$0.73$ &   &  \phs$0.15$ &  \phs$0.65$
\enddata
\tablenotetext{a}{Spearman rank correlation coefficients and the probability that the variables are \emph{uncorrelated}.  We compare the residuals, $\Delta\log({\rm O/H})$, for all three integrated abundance measurements (see \S\ref{sec:intoh} for details), against the slope of the abundance gradient, the dust reddening, $E(B-V)$, the inclination angle, $i$, and the H$\alpha$ surface brightness, $\Sigma({\rm H}\alpha)$.}
\tablenotetext{b}{Oxygen abundances derived using either the \citet[M91]{mcgaugh91} or the \citet[PT05]{pilyugin05} abundance calibrations.  See \S\ref{sec:intoh} for details.}
\end{deluxetable}

\end{document}